\documentclass[twocolumn, trackchanges]{aastex63}

\newcommand{\dif}{\mathrm{d}}

\usepackage{amsmath}
\usepackage{lineno}
\usepackage{booktabs}
\usepackage{float}

\shorttitle{Tilting Uranus}
\shortauthors{Lu \& Laughlin}

\begin{document}

\title{Tilting Uranus via Spin-Orbit Resonance with Planet Nine}

\author[0000-0003-0834-8645]{Tiger Lu}
\affiliation{Yale University, 52 Hillhouse, New Haven, CT 06511, USA}

\author[0000-0002-3253-2621]{Gregory Laughlin}
\affiliation{Yale University, 52 Hillhouse, New Haven, CT 06511, USA}

\received{2021 November 1}
\revised{2022 March 2}
\accepted{2022 July 22}
\submitjournal{PSJ}

\begin{abstract}
Uranus’ startlingly large obliquity of $98^\circ$ has yet to admit a satisfactory explanation. The most widely accepted hypothesis involving a giant impactor that tipped Uranus onto its side encounters several difficulties with regards to the Uranus' spin rate and its prograde satellite system. An obliquity increase that was driven by capture of Uranus into a secular spin-orbit resonance remains a possible alternative hypothesis that avoids many of the issues associated with a giant impact. We propose that secular spin-orbit resonance could have excited Uranus’ obliquity to its present day value if it was driven by the outward migration of an as-yet undetected outer Solar System body commonly known as Planet Nine. We draw support for our hypothesis from an analysis of 123 N-body simulations with varying parameters for Planet Nine and its migration. We find that in multiple instances, a simulated Planet Nine drives Uranus' obliquity past $98^\circ$, with a significant number falling within $10\%$ of this value. We note a significant caveat to our results in that a much faster than present-day spin-axis precession rate for Uranus is required in all cases for it to reach high obliquities. We conclude that while it was in principle possible for Planet Nine (if it exists) to have been responsible for Uranus’ obliquity, the feasibility of such a result hinges on Uranus' primordial precession rate.
\end{abstract}


\section{\textbf{Introduction}} 
\label{sec:intro}
The large obliquity of Uranus is a puzzle. Conservation of angular momentum in the primordial Solar System naively suggests that as primordial gas giant planets accrete from the planetary disk, their axial tilts should be driven to zero. In reality, however, the Solar System's gas giants span a range of obliquities, with Uranus being most extreme with $\theta=98.7^\circ$. The prevailing and best-studied theory posits that some type of giant impact was responsible - likely a 1-3 $M_\Earth$ body impacting the primordial Uranus, simultaneously generating the large axial tilt and spurring the formation of its satellite system \citep{harris1982dynamical, benz_1989, SLATTERY1992, izidoro2015accretion, Kegerreis_2018, kegerreis2019, Reinhardt_2019, ida_2020, rogoszinski2021tilting}. The theory has significant merit and giant impacts are indeed both feasible and capable of explaining much of Uranus' present-day configuration. In a few aspects, however, it encounters difficulties. \cite{rogoszinski2021tilting} present an in-depth discussion of both the merits and drawbacks of the giant impact hypothesis. An immediate issue lies in the near-match between the spin rates of Uranus and Neptune. Both Uranus and Neptune spin significantly slower (with $T_U \sim 17.2$ hours and $T_N \sim 16.1$ hours) than their break-up speeds, which would naively be expected for an accreting gas giant \citep{machida_2008, batygin2018terminal, bryan2018constraints, dong_2021, dittman_2021}. A giant impact would likely change a planet's spin rate - however, \cite{rogoszinski2021tilting} show that the impactors required to explain both Uranus' and Neptune's obliquities $(\theta_N \sim 30^\circ)$ are unlikely to spin both planets down similarly. In addition, the prograde Uranian satellite system at first glance appears incompatible with a giant impact. \cite{morbi2012} argues that given a single impactor the system would be expected to have a retrograde orbit. To account for the observed prograde orbit, Uranus would either require an initially large obliquity (on the order of $\sim 10^\circ$) or multiple impactors. Furthermore, impactors that are sufficiently large to explain the axial tilt of Uranus (greater than 1 $M_\Earth$) are expected to produce disks that exceed the total mass of the Uranian satellite system by an order of magnitude \citep{Kegerreis_2018, Reinhardt_2019}. A possible solution has been offered by \cite{ida_2020}, who show that the incongruity in impact-generated disks can be reconciled if the evolution of a water-vapor disk is taken into account. By assuming ice-dominated compositions for both Uranus and its impactor, Uranus' present-day satellite system is reproduced.

We investigate an alternative explanation of Uranus' large axial tilt by positing the capture and subsequent evolution of the planet in secular spin-orbit resonance. The resonant capture mechanism is slow enough to preserve both the structure of the planet's satellite system and its spin rate \citep{goldreich1965}, which circumvents many of the problems with the giant impact theory. Furthermore, the process is potentially responsible for Jupiter's $3^\circ$ \citep{Ward_2006,vokrouhlicky2015tilting} and Saturn's $27^\circ$ \citep{ward2004, hamilton2004} obliquities; plausibility in these instances is supported by the near-match between Jupiter's and Saturn's spin precession frequencies and the contibuting nodal regression frequencies driven by Uranus and Neptune, respectively. Recently, however, \cite{saillenfest2021large, saillenfest2021past} have argued that early resonance with Neptune is incompatible with the fast tidal migration of Titan, and suggest that a later entry to the resonance driven by nodal precession evolution stemming from Titan's migration is responsible for Saturn's obliquity. A similar process, driven by the migration of the Galilean satellites, may excite Jupiter's currently small obliquity to larger values in the future \citep{saillenfest2020future}.

Uranus' current spin precession frequency is far too slow to be a near-match for any mode of secular forcing provided by the present-day Solar System. Significant work has thus been done to explore past Uranian evolutionary pathways which would result in a closer match. \cite{millholland_2019} studied the effect of an evolving circumstellar disk on the obliquities of both Uranus and Neptune, and concluded that disk-induced spin-orbit resonance is unlikely to be the cause of their axial tilts. \cite{Boue_2010} posit that Uranus' current obliquity is possible if Uranus had an additional large moon present in the past - however, this moon would require a mass of up to $1 \%$ the mass of Uranus and would have needed to be dispensed with after the resonance's action was complete. The study of \cite{rogoszinski2020tilting} posited a circumplanetary disk that increased the Uranian spin precession frequency, and was able to account for obliquities of up to $70^\circ$. More recently, \cite{rogoszinski2021tilting} investigated the effect of Neptune's migration on a primordial Uranus placed between Jupiter and Saturn. They concluded that $90^\circ$ obliquities are achievable, but only on unrealistic timescales; $40^\circ$ tilts are obtainable upon more reasonable timescales. In all cases, however, impacts must be invoked to subsequently drive the obliquity to the present-day value of $98^\circ$.

The review of \cite{Batygin_2019} summarizes the development of the hypothesis that the observed orbital alignment of long-period trans-Neptunian objects is maintained by the presence of an as-yet undetected (5 - 10 $M_\Earth)$ body (known as Planet Nine) in the outer reaches of the Solar System. The dynamical effect that such a body, if present, would exert on Uranus thus merits investigation. In this article, we examine the effect that dynamical evolution (in the form of outward migration) of Planet Nine would have on Uranian spin axis, and we find that under certain conditions, a Planet Nine could have generated the dramatic spin-orbit misalignment that constitutes Uranus' most uniquely defining feature.

The paper is structured as follows. In Section \ref{sec:pf}, we outline the relevant values of axial precession and nodal regression. In Section \ref{sec:resonance}, we use these quantities to discuss the physics involved in the process of secular spin-orbit resonance, and introduce the equation of motion that dictates the evolution of the spin axis. In Section \ref{sec:node}, we describe our procedure for modeling the evolution of the Uranian node. We also discuss how we model the dynamical evolution of Planet Nine and its effect on the Uranian orbital evolution. In Section \ref{sec:nbody}, we report the resulting obliquities we are able to achieve, and we discuss and draw conclusions in Section \ref{sec:disc}.

\section{\textbf{Precessional Frequencies}} \label{sec:pf}
\begin{figure}
    \centering
    \includegraphics[width=0.4\textwidth]{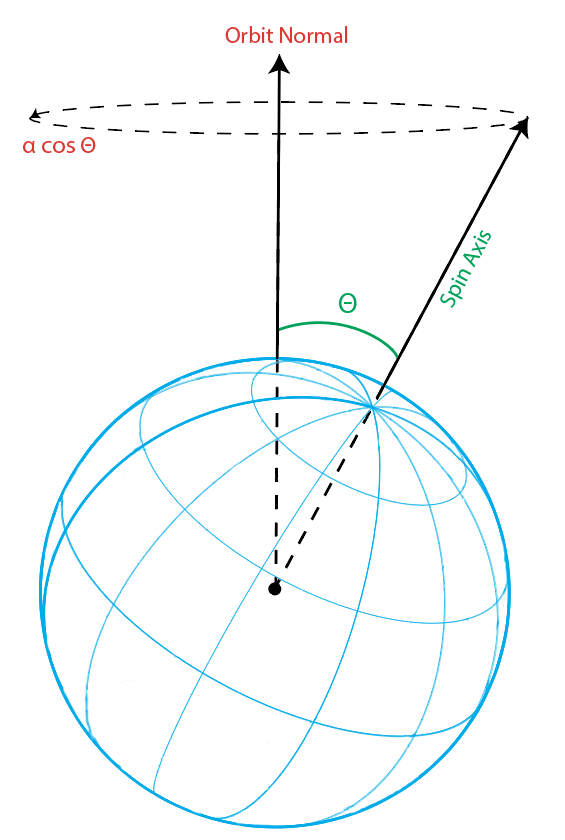}
    \caption{Torques associated with the host star and satellite system cause the spin axis to precess about the orbit normal with frequency $\alpha \cos \theta$.}
    \label{fig:1}
\end{figure}

In the presence of torques from a host star and a satellite system, a planet's spin axis will precess about its orbit normal \citep{goldreich1965} with period (given low orbital eccentricity)

\begin{equation}
    T_\alpha = \frac{2 \pi}{\left| \alpha \cos \theta \right|},
    \label{eq:tau_a}
\end{equation}

\noindent where $\theta$ is defined as the planet's obliquity, or the angle between the spin axis and orbit normal, and $\alpha$ is defined as the \textit{precessional constant}. Figure \ref{fig:1} provides a schematic representation.

The precessional constant parameterizes the rate at which the spin axis precesses about the orbit normal. In the case where the satellite system exhibits prograde orbits in the equatorial plane (as is the case for the Uranian system), it is given by \citep{TREMAINE1991, french1993geometry, rogoszinski2021tilting} 

\begin{equation}
    \alpha = \frac{3 n^2}{2 \omega} \frac{J_2 + q}{\lambda + l},
\label{eq:alpha}
\end{equation}

\noindent where $\omega$ is the planet's spin frequency, $n$ its orbital mean motion about the host star, $J_2$ is the quadrupole moment of the planet's gravitational field, and $\lambda$ is its moment of inertia normalized by $M_p R_p^2$. The quantities $q$ and $l$ account for elements of the planet's satellite system or circumplanetary disk, with $q$ being the effective quadrupole coefficient and $l$ the angular momentum normalized by $M_p R_p^2 \omega$. For a satellite system \citep{TREMAINE1991, french1993geometry, ward2004, rogoszinski2021tilting} one has

\begin{equation}
\begin{split}
    & q \equiv \frac{1}{2} \sum_i \left(\frac{M_i}{M_p}\right) \left(\frac{a_i}{R_P}\right)^2\, \\
    & l \equiv \frac{1}{M_p R_p^2 \omega}\sum_i  M_i a_i^{2} n_i,
\end{split}
\label{eq:ql_moon}
\end{equation}

with the sum running over all $i$ satellites, and where $M_i, a_i, n_i$ are the mass, semi-major axis and mean motion, respectively, of each satellite.

\noindent Considering the major satellites of Uranus, from Equations \ref{eq:alpha} and \ref{eq:ql_moon}, we can calculate the present-day value of $\alpha$ for Uranus. Using $J_2 = 3.343 \times 10^{-3}$ \citep{dermott_1984} and $\lambda = 0.225$ \citep{yoder1995astrometric}, we find that $\alpha = 0.045$ arcseconds/yr. At its current $\theta = 98^\circ$ obliquity, Uranus' axial precession period is $T_\alpha = 169$ Myr.

\begin{figure}
    \centering
    \includegraphics[width=0.45\textwidth]{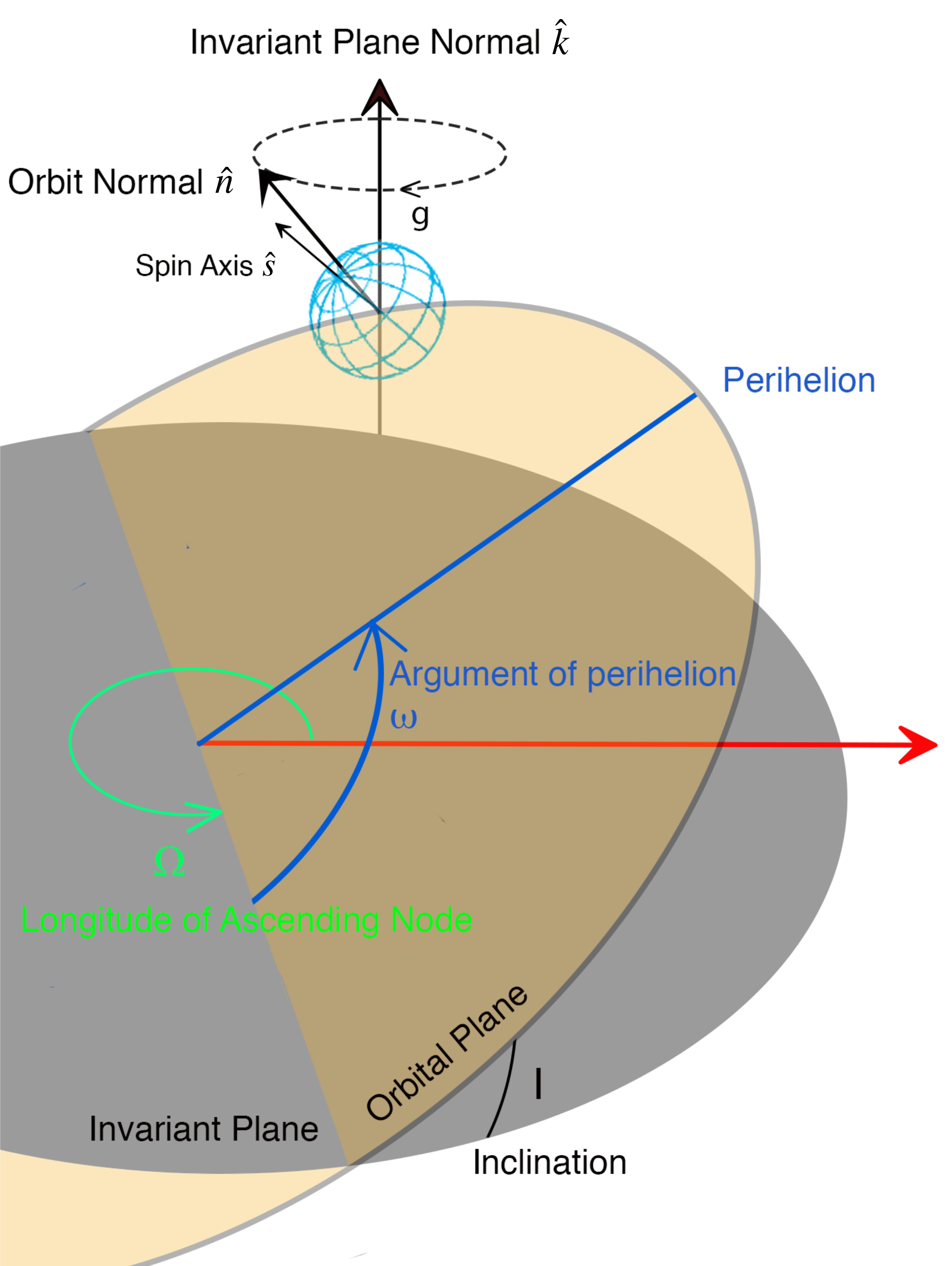}
    \caption{Torques associated with perturbing planets cause the inclined orbit of a planet (with normal $\hat{n}$) to recess about the invariant plane normal $\hat{k}$, at a rate of $g = \dot{\Omega}$. The planet's spin axis $\hat{s}$ is also shown. Not shown on this figure is the longitude of periapsis $\varpi$, defined as $\varpi = \Omega + \omega$}
    \label{fig:2}
\end{figure}

Simultaneously, in the presence of torques from neighboring planets and other large bodies, the orbital plane of a planet will recess about the normal to the invariant plane of the system (the plane perpendicular to the system's overall angular momentum vector). The period of recession is given by $T_g = 2 \pi/|g|$, with $g = \dot{\Omega}$, where $\Omega$ is the longitude of the ascending node. Figure \ref{fig:2} shows a schematic view of the motion.

\section{\textbf{Spin-Orbit Resonance}} \label{sec:resonance}
Secular spin-orbit resonance trapping has repeatedly been shown to be a possible mechanism for generating substantial planetary obliquities. The dynamics of the phenomenon are well-studied \citep{peale_1969}, and its action has been shown to be plausibly responsible for Jupiter's $3^\circ$ \citep{Ward_2006} and Saturn's $26.7^\circ$ \citep{ward2004, hamilton2004, saillenfest2021past, saillenfest2021large} axial tilts. Exciting obliquities via secular spin-orbit resonance has been shown to be self-consistent with the orbital migrations of Uranus and Neptune predicted by the Nice Model \citep{vokrouhlicky2015tilting}.

\subsection{First Order Spin Axis Equation of Motion}
To a first order of approximation, resonant motion of a planet's spin axis can be exhibited with the restrictive assumption of constant $I$, $\alpha$ and $g$. In a coordinate frame rotating with angular frequency $g$ centered on the orbit normal (which we will call the ``rotating frame"), the equation of motion can be approximated \citep{TREMAINE1991, ward2004}

\begin{equation}
    \frac{\dif \hat{s}^*}{\dif t} = \alpha \left( \hat{s}^* \cdot \hat{n} \right) ( \hat{s}^* \times \hat{n} ) + g ( \hat{s}^* \times \hat{k} ),
    \label{eq:sim_eom}
\end{equation}

\noindent where $\hat{s}^*, \hat{n}, \hat{k}$ are the unit vectors pointing in the direction of the spin axis, orbit normal, and invariant plane normal, respectively. The superscript $^*$ notation on the unit spin vector is to differentiate the spin vector defined here in the rotating frame as opposed to in the invariant plane of the system, which will be defined later.

\subsection{Cassini States}
The spin axis may be defined in terms of two angles - the obliquity $\theta$, previously defined, and the \textit{precession angle} $\psi$, which is defined as the angle between the projections of $\hat{s}$ and $\hat{k}$ onto the plane perpendicular to $\hat{n}$. These angles can be related to the position of the unit spin vector in Cartesian coordinates \cite{morbidelli2002modern}

\begin{equation}
    \hat{s_x} = \sin \theta \sin \psi, \hat{s_y} = \sin \theta \cos \psi, \hat{s_z} = \cos \theta
\end{equation}

\noindent The Hamiltonian that governs the motion of the spin axis is well known \citep{morbidelli2002modern}. In Cartesian coordinates (in the rotating frame),

\begin{equation}
\begin{split}
    \mathcal{H} & = \frac{\alpha}{2} \left(1 - \frac{1}{2}x^2 - \frac{1}{2}y^2 \right)^2 + \frac{\cos(I)}{2} g \left( x^2 + y^2 \right) \\
    & - \frac{\sin(I)}{2} g x \sqrt{4 - x^2 - y^2},
\end{split}
\end{equation}

\noindent where $I$ is the planet's inclination, and the Cartesian coordinates $x$ and $y$ can be expressed in terms of the obliquity and precession angle

\begin{equation}
    x = 2 \sin \frac{\theta}{2} \cos \psi, y = 2 \sin \frac{\theta}{2} \sin \psi
\end{equation}

\begin{figure}
    \centering
    \includegraphics[width=0.45\textwidth]{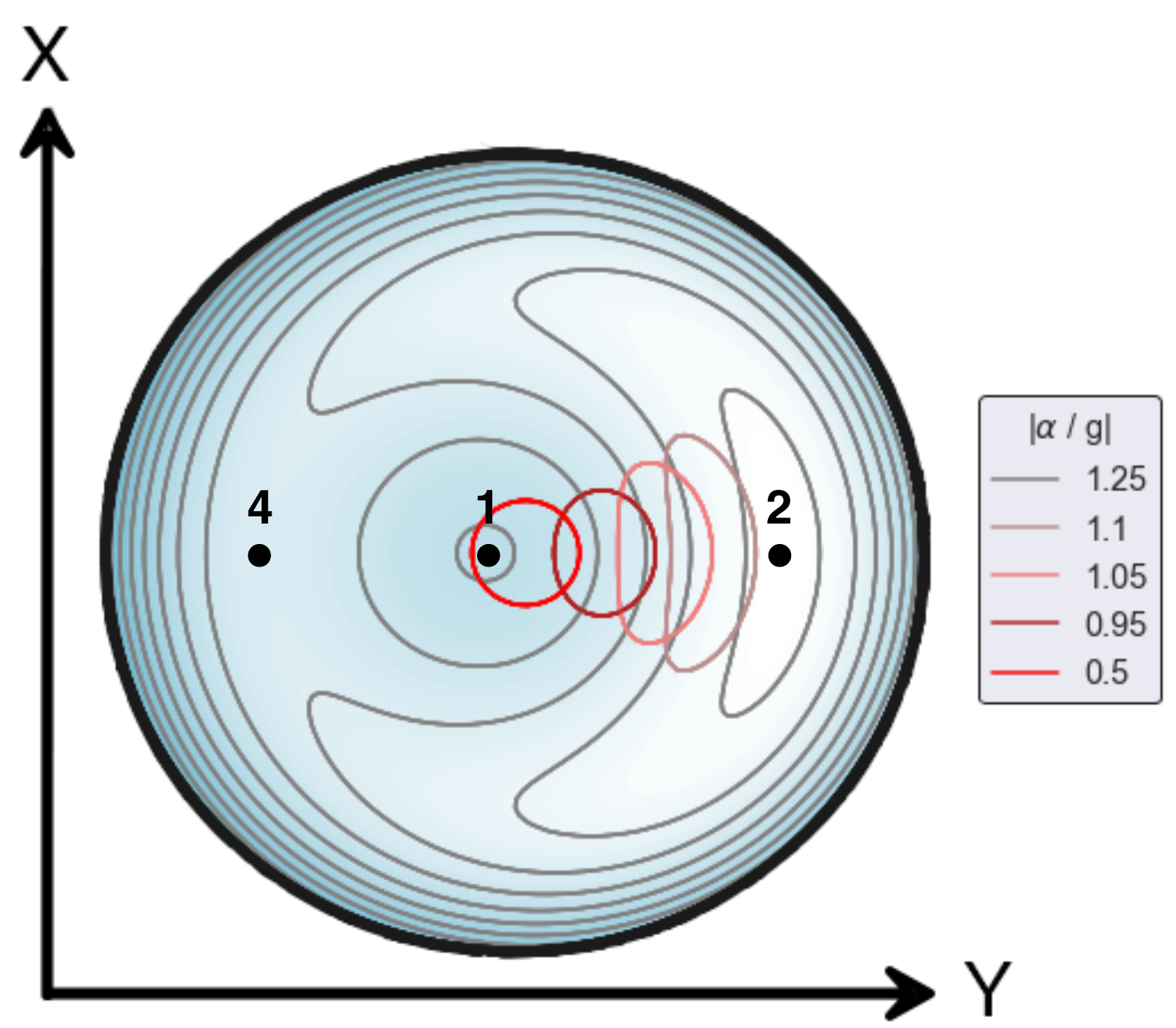}
    \caption{Level curves of the spin axis Hamiltonian (in black) with $\alpha / g = 1.25 > (\alpha / g)_{crit}$, with the Cassini state equilibrium points labelled (Cassini State 3 corresponds to a retrograde spin state and is not shown). Spin axis trajectories are confined to these level curves. The colored lines tracks the evolution of one such trajectory as the ratio $|\alpha / g|$ changes - note how the curve begins as a circulating trajectory and evolves into a librating one.}
    \label{fig:3}
\end{figure}

For a full derivation, see \cite{millholland_2019}. The motion of the spin vector is confined to the level curves (curves of constant energy) of the Hamiltonian. As the ratio $\alpha / g$ evolves, so do the locations of the level curves. At a critical value,

\begin{equation}
    (\alpha / g)_{crit} = - \left( \sin^{2/3} I + \cos^{2/3} I \right)^{3/2},
\end{equation}

\noindent a separatrix appears and additional equilibrium points emerge \citep{ward2004, fabryky2007}. See Figure \ref{fig:3} for a plot of the level curves of the Hamiltonian in the rotating frame of the planet at $\alpha / g > (\alpha / g)_{crit}$ (the necessary criterion for resonant capture, see Section \ref{sec:rescap}). The equilibrium points at the extrema of the Hamiltonian are known as \textit{Cassini states}, and correspond to configurations such that in the invariant frame, \citep{1966_colombo, peale_1969}

\begin{itemize}
    \item $\hat{s}, \hat{n}$ and $\hat{k}$ are coplanar
    \item $\hat{s}$ and $\hat{n}$ precess about $\hat{k}$ at the same rate.
\end{itemize}

\noindent Clearly, in the rotating frame centered on the orbit normal, the spin axis position appears stationary. Four such equilibrium points exist: states $1$, $2$ and $4$ are shown on Figure \ref{fig:3}, while state $3$ is retrograde and thus projects into the downward-facing hemisphere. For a pair of values of $\alpha / g$ and inclination $I$, the obliquities $\theta_c$ of the corresponding Cassini states are given by \citep{ward2004}

\begin{equation}
    \frac{\alpha}{g} \cos \theta_c \sin \theta_c + \sin \left(\theta_c - I \right) = 0.
\end{equation}

\noindent For $\alpha / g < (\alpha / g)_{crit}$ states 1 and 4 do not exist \citep{ward2004}, so we consider motions of the spin axis about Cassini State 2. Figure \ref{fig:3} shows that for trajectories that lie close to Cassini state 2, the spin axis will librate in a banana-like trajectory about the equilibrium, rather than circulating the origin.

\subsection{Resonant Capture}
\label{sec:rescap}

If the ratio $\alpha / g$ changes slowly enough, the phase space area enclosed by a trajectory remains constant. This is known as the \textit{adiabatic criterion} and is satisfied if the rate of change in $\alpha / g$ is much slower than a libration timescale \citep{millholland_2019}. As the ratio $|\alpha / g|$ increases, Cassini state 2 will move outward and an orbital trajectory trapped in resonance will be forced to larger and larger obliquities. This evolution is overplotted in Figure \ref{fig:3}.

\begin{figure*}
    \centering
    \includegraphics[width = 1\textwidth]{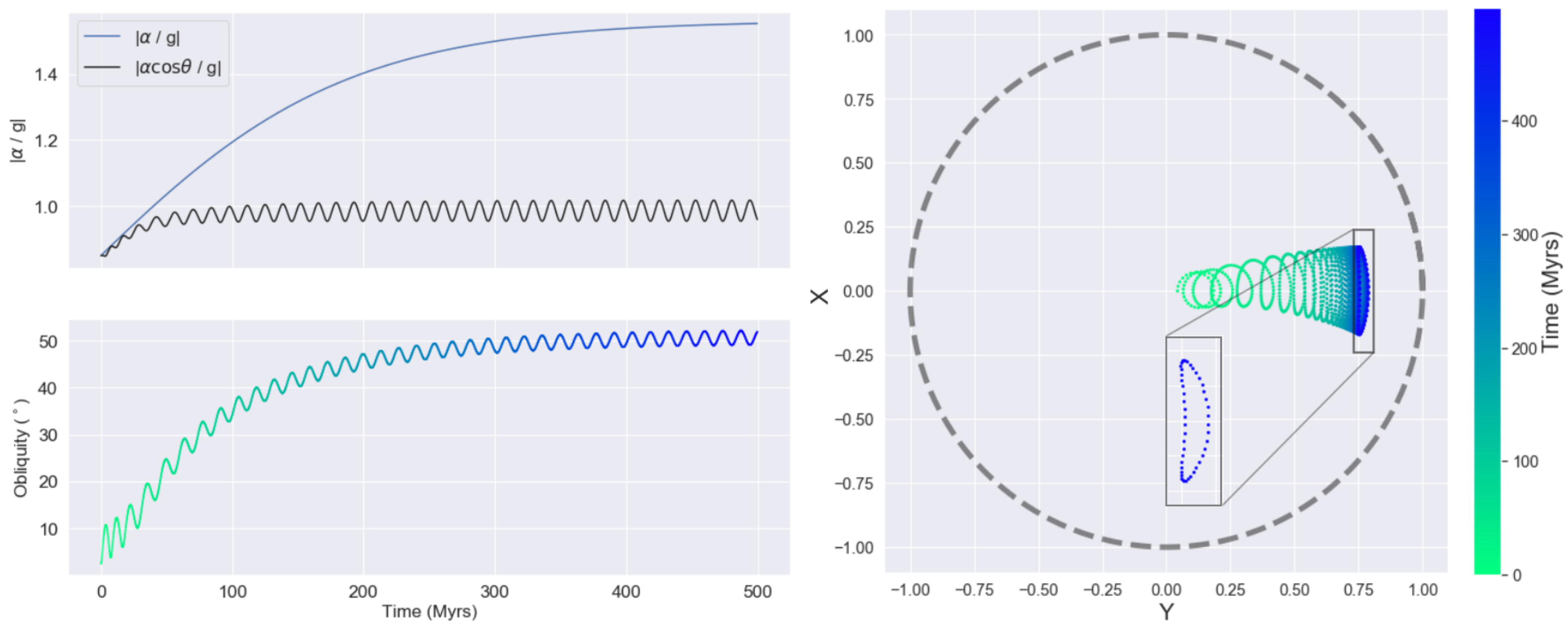}
    \caption{Result of a simple evolution of the ratio $|\alpha / g|$, obtained via integration of Equation \ref{eq:sim_eom}. The upper left plot displays the evolution of $|\alpha / g|$ imposed on this simulation - note that the ratio $|\alpha \cos \theta / g|$ saturates at unity. The bottom left plot shows the resulting obliquity evolution. The right plot shows the location of the spin axis over the evolution period. We have zoomed in on one libration timescale to explicitly show the banana-like libration motion the spin axis exhibits when captured into resonance.}
    \label{fig:evolution}
\end{figure*}

If the adiabatic condition is satisfied, resonant capture occurs when the ratio $|\alpha \cos \theta / g|$ approaches unity from \textit{below}. At low obliquities, this is equivalent to the condition that $|\alpha / g|$ approach unity from below. By contrast, a passage through unity from the opposite direction leads to an impulsive, potentially large kick to the obliquity, without capture into resonance \citep{ward2004}. If resonance capture occurs, the ratio $T_g / T_\alpha$ oscillates around an average value of unity and large increases in obliquity can occur. Figure \ref{fig:evolution} shows an example evolution of the ratio $|\alpha / g|$ and the resultant obliquity evolution.

\subsection{Spin Axis Equation of Motion}
In reality, a planet's orbital inclination is not constant, the orbit is eccentric, and the value of $\alpha$ can vary with time. The simple equation of motion in this event no longer gives an accurate picture. In the invariant plane of the solar system, the orbital inclination and longitude of ascending node may be decomposed into more than one harmonic. The motion of the unit spin axis $\hat{s}$ in this regime is given by \citep{ward_1979, hamilton2004}:

\begin{equation}
\label{eq:spin_eom}
    \begin{bmatrix}
    \dot{s_x} \\ \dot{s_y} \\ \dot{s_z}
    \end{bmatrix}
    = \frac{\alpha \left( s_x p \xi - s_y q \xi + s_z \eta \right)}{(1 - e^2)^{3/2}}
    \begin{bmatrix}
    s_y \eta + s_z q \xi \\
    s_z p \xi - s_x \eta \\
    -s_x q \xi - s_y p \xi
    \end{bmatrix},
\end{equation}

\noindent which in turn draws on functions of the planet's orbital elements

\begin{equation}
    \begin{split}
        p & = 2 \sin \left(\frac{I}{2} \right) \sin \Omega \\
        q & = 2 \sin \left(\frac{I}{2} \right) \cos \Omega \\
        \xi & \equiv \sqrt{1 - \frac{1}{4}(p^2 + q^2)} \\
        \eta & \equiv 1 - \frac{1}{2}(p^2 + q^2)\, ,
    \end{split}
    \label{eq:elements}
\end{equation}

\noindent where $e$ is eccentricity, $I$ is inclination, $\varpi$ is the longitude of periapsis and $\Omega$ the longitude of ascending node. To see librations we must transform back into the frame of the planet's orbit via \citep{ward_1974}

\begin{equation}
    \hat{s}^* = A \hat{s},
\end{equation}

\noindent where $\hat{s}^*$ is the unit spin axis in the frame that rotates with the angular momentum vector of the planetary orbit, $\hat{s}$ is the unit spin axis in the invariant plane, and $A$ is the time-dependent rotation matrix

\begin{equation}
    A = \begin{bmatrix}
    \cos \Omega & \sin \Omega & 0 \\
    - \cos I \sin \Omega & \cos I \cos \Omega & \sin I \\
    \sin I \sin \Omega & - \sin I \cos \Omega & \cos I
    \end{bmatrix},
\end{equation}

\noindent where $\Omega(t)$ and $I(t)$ vary in time.

\section{\textbf{Nodal Evolution Modeling}} \label{sec:node}
Approaches of escalating complexity can be employed to model the evolution of the Uranian nodal precession, $g$ in response to perturbations from the other bodies of the Solar System.

\subsection{Laplace-Lagrange Secular Theory}

Short of full numerical simulations that track the Uranian node, significant physical intuition can be gained from the linear approximation provided by Laplace-Lagrange secular theory. Secular theory describes the orbital motion of planetary orbits as an averaged approximation of their long term motions. Terms in the gravitational disturbing function that include the mean longitude are ignored in this approximation, as they vary relatively rapidly and in the long term average to zero. In the late 1700s, it was determined that to lowest order, the time dependence of the eccentricities $e$ \citep{laplace_1772} and the inclinations $I$ \citep{lagrange_1778} of a planetary system with $k$ bodies are described by a system of first-order linear differential equations

\begin{equation}
    \frac{\dif}{\dif t}
    \begin{bmatrix}
    z_1 \\
    \vdots \\
    z_k \\
    \zeta_1 \\
    \vdots \\
    \zeta_k \\
    \end{bmatrix}
    =
    \sqrt{-1} \begin{bmatrix}
    A_k & 0_k \\
    0_k & B_k
    \end{bmatrix}
    \begin{bmatrix}
    z_1 \\
    \vdots \\
    z_k \\
    \zeta_1 \\
    \vdots \\
    \zeta_k \\
    \end{bmatrix}.
\end{equation}

\noindent Which was applied to the motion of the solar system planets by \cite{laplace_1784} (for an in-depth review of the development of Laplace-Lagrange secular theory, see \citep{laskar2013solar}). Here $z \equiv e \exp \sqrt{-1} \varpi$ and $\zeta \equiv 2 \sin (I / 2) \exp \sqrt{-1} \Omega$. $A_k$ and $B_k$ are $k \times k$ matrices whose elements depend solely on the masses and semi-major axes of the $k$ planets, and $0_k$ denotes the $k \times k$ zero matrix. At this level of approximation, the time evolution of $z_i$ and $\zeta_i$ is given as a sum of sinusoidal contributions

\begin{equation}
    \begin{split}
    z_i & = \sum_{j = 1}^k \alpha_{ij} e^{\sqrt{-1} \gamma_j t} \\
    \zeta_i & = \sum_{j = 1}^k \beta_{ij} e^{\sqrt{-1} \delta_j t}, \\
    \end{split}
\end{equation}

\noindent where the quantities $\delta_j$, $\alpha_{ij}$, $\gamma_j$, $\beta_{ij}$ are the eigenvalues and eigenvectors of the $A_k$ and $B_k$ matrices, respectively. The $i$ and $j$ indices range from 1 to $k$ and represent each planet in the system. Of particular interest to our analysis is the quantity $\delta_j$, which determines the ``forcing frequency'' at which the $j$th body perturbs $\Omega$.

A slightly modified version of the foregoing theory that is specific to the Solar System (and which partially accounts for additional Jupiter-Saturn interactions) was developed by \cite{bvw_1960}, and was notably reprinted in \citep{murray_dermott_2000}.

\begin{figure}
    \centering
    \includegraphics[width= 0.45\textwidth]{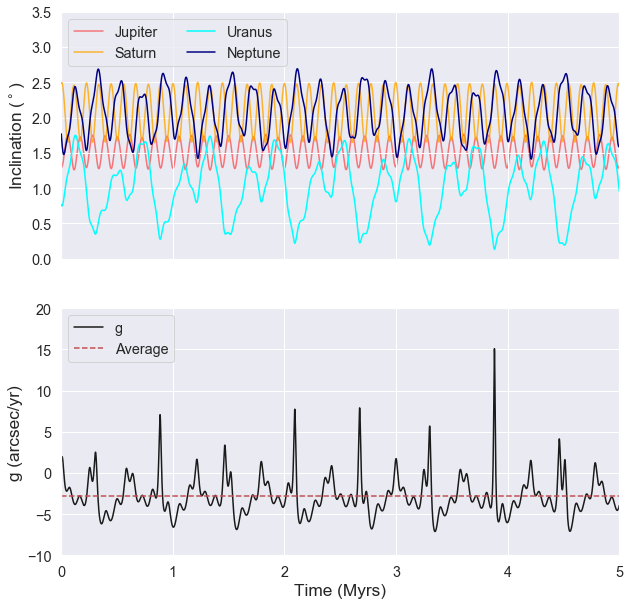}
    \caption{The evolution of the four outer planets as described by the secular theory of \cite{laplace_1784}. The upper plot shows the evolution of inclination of all four outer planets - note the complex yet semi-periodic behavior. The bottom plot shows the evolution of $g$, the rate of change of Uranus's node position.}
    \label{fig:secular}
\end{figure}

Given the time evolution of $z_i$ and $\zeta_i$, the orbital elements (see Equation \ref{eq:elements}) including $\Omega$ can be recovered. Figure \ref{fig:secular} shows the inclination evolution of the four outer planets implied by secular theory, as well as the evolution of the Uranian $g$.

\subsection{Effect of Planet Nine}

Planet Nine is a hypothesized body in the outer Solar System whose existence has been inferred from the apparent apsidal confinement of long-period trans-Neptunian objects \citep{Batygin_2016}. For a more detailed review of the evidence for Planet Nine, see \citep{Batygin_2019}. The most recent Markov Chain Monte Carlo simulations \citep{brown2021orbit} give Planet Nine's best-fit orbital parameters:

\begin{equation}
    \begin{split}
        m_9 & = 6.9^{+2.6}_{-1.6} \text{ M}_\Earth\\
        a_9 & = 460.7^{+178.8}_{-103.3} \text{ AU} \\
        e_9 & = 0.3^{+0.1}_{-0.1} \\
        i_9 & = 15.6^{+5.2}_{-5.4} \text{ }^\circ. \\
    \end{split}
    \label{eq:p9params}
\end{equation}

\noindent In contrast, however, the recent study by \cite{Batygin_2021} suggests that a more eccentric and more distant Planet Nine may be required.

\textit{In situ} formation of Planet Nine is believed to be unlikely \citep{kenyon2016making}, and discussion has settled on two preferred possibilities for its origin. One possibility is that it formed alongside the four giant planets and subsequently migrated outward to a much more distant orbit \citep{thommes1999formation, brasser2012reassessing, izidoro2015accretion, Bromley_2016, li2016interaction, Eriksson_2018}. A second possibility is that it was captured from a passing star within the Solar System's stellar birth aggregate \citep{li2016interaction, mustill2016there, parker2017planet}. If this second scenario holds, there would be no plausible connection to the Uranian obliquity, so we focus on long-distance outward migration of Planet Nine as an assumed formation pathway. The parameters given in Equation \ref{eq:p9params} assume a prior involving Planet Nine's origin and eventual ejection from the Jupiter-Saturn region.

\cite{Batygin_2019} proposed a two-step migration process: First, Jupiter or Saturn scatter Planet Nine onto a temporary high-eccentricity orbit, which was then circularized via gravitational perturbations from nearby stars in the cluster. It should be stressed, however, that this is an unlikely scenario - the probability of a Planet Nine-sized body settling between 100 and 5000 AU due to stellar perturbations is no more than a few percent \citep{Bailey_2019}. A more likely alternative involves circularization via dynamical friction with a circumstellar disk. With a gaseous disk, \cite{Bromley_2016} were able to reproduce orbits similar to that of Planet Nine post-scattering, with a preference towards very slowly decaying disks (with dissipation timescales on the order of 10 Myr). Subsequently, numerical simulations of \cite{Carrera_2017} were used to suggest that a 60-130 M$_\Earth$ planetesimal disk formed beyond 100 AU as a consequence of the streaming instability. \cite{Eriksson_2018} found that a $10$ M$_\Earth$ planet scattered into such a disk from the vicinity of Neptune's orbit has a $20-30\%$ chance of reproducing a Planet Nine-like orbit.

\begin{figure}
    \centering
    \includegraphics[width = 0.45\textwidth]{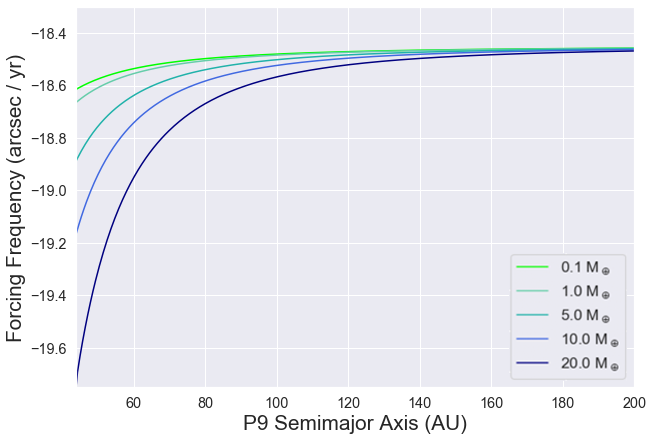}
    \caption{Planet Nine's forcing frequency on Uranus as a function of semi-major axis. Each line corresponds to a Planet Nine of varying mass (in Earth masses).Note that the forcing frequency does not go to zero even as Planet Nine's mass does - this is because each eigenfrequency does not correspond one-to-one with a planet, but rather is simply dominated by a given body.}
    \label{fig:forcing}
\end{figure}

As Planet Nine moves outward in its putative migratory trajectory, the magnitude of its perturbations on the Uranian $\Omega$ steadily diminish (see Figure \ref{fig:forcing}). The rate of change of the Uranian node, $g$, thus decreases with time. In the event that the precession factor, $\alpha$, remains constant, this provides a pathway for resonance capture as the ratio $|\alpha / g|$ evolves to reach unity.

\subsection{Numerical Simulations}
The Laplace-Lagrange theory assumes constant semi-major axes for all planets, and additionally fails in the limit of significant short-period terms \citep{Camargo_2018}, so a migrating Planet Nine cannot be modelled in this framework. We instead turn to full N-body integrations that employ the REBOUND package with the adaptive high-order IAS15 integrator \citep{ias15}. We incorporated orbit-averaged physical forces to simulate the outward migration of Planet Nine using the \textit{modify\_orbits\_forces} implementation \citep{kostov_2016} in the REBOUNDx package \citep{Tamayo_2019}.

An overview of our modeling procedure is as follows. A REBOUND simulation is initialized with the Sun and the four outer planets in their present-day configuration, with orbital element values from the NASA HORIZONS database. Planet Nine is then initialized in its starting position. Free parameters associated with Planet Nine include its mass, its eccentricity, and its inclination. We set the initial semi-major axis and $\tau_a$ (which parameterizes the migration rate as seen in \cite{kostov_2016}) constant. The system as thus specified is in the frame of the ecliptic. We then determine the invariant plane of the Solar System with Planet Nine accounted for, via the method outlined by \cite{suoami}. The normal vector to the invariant plane is

\begin{equation}
    \hat{L}_{tot} = \sum_{j = 1}^N m_j \hat{r}_j \times \hat{v_j},
\end{equation}

\noindent where $m_j, r_j, v_j$ are the mass, barycentric position, and barycentric velocity of the $j$th body, respectively. We recalculate the inclination of each planet in the invariant frame by taking the angle between $\hat{L}_{tot}$ and the planet's own orbital angular momentum vector. Each simulation is then integrated forward for $10^8$ years. The results of the REBOUND simulation (namely, values for the Uranian inclination, eccentricity, $\Omega$, and $\varpi$) are used to construct time evolutions of the orbital elements $e(t)$, $p(t), q(t)$. The orbital elements at each timestep are then passed into the spin axis equation of motion (eq. \ref{eq:spin_eom}) to track the motion of the spin axis. The rate of nodal regression $g = \dot{\Omega}$ is itself not used as an input to our spin vector equation of motion, but is useful for illustrative purposes - this quantity is calculated numerically from the value of $\Omega(t)$ recovered at each timestep. In each case, Uranus' spin axis is initialized with an obliquity of $2.5^\circ$. This value is arbitrarily chosen, but any small initial tilt should yield similar results. We test each REBOUND simulation with various values of $\alpha$ linearly spaced between $0.005$ and $6$ arcseconds / yr and track the results.

\section{\textbf{Results of N-Body Simulations}} \label{sec:nbody}

In this section we present selected results from our N-body simulations. For a compiled list of all our simulation results as well as the source code used, see \url{www.github.com/tigerchenlu98/tilting-uranus}. Throughout all of our simulations, the magnitude of the spin vector differs from unity by less than one part in $10^5$.

\subsection{Solar System Model}
\label{sec:ssm}

We now examine simulations of a Solar System dynamical model that includes Jupiter, Saturn, Uranus, Neptune, and Planet Nine. To save computation time the terrestrial planets were not included, as their effects are negligible. For each simulation, we initialize Planet Nine at $40$ AU. The $\tau_a$ parameter, as seen in \cite{kostov_2016}, parameterizes the rate of change in the system. The evolution of Planet Nine's semi-major axis is given by

\begin{equation}
    a = a_0 e^{t / \tau_a},
\end{equation}

\noindent In practice, perturbative influences from the other planets lead to slightly more complex realized semi-major axis evolution and hence a non-deterministic final $a$. We set $\tau_a = 4 \times 10^7$ years for each of our simulations - for this value, the semi-major axis progresses through 2.5 e-folding timescales. \noindent We emphasize that our choice of initial $a$ and $\tau_a$ do not reflect Planet Nine's true initial position or migratory trajectory - recall that Planet Nine is believed to have originated with the giant planets. Rather, the value of 40 AU is picked to encapsulate as much of the evolutionary trajectory as possible, without sacrificing the stability of the simulation. As the effect of Planet Nine on the Uranian $\Omega$ only increases for smaller semi-major axis (per secular theory), bringing Planet Nine's initial position closer in would serve to bring the ratio $\alpha / g$ down, which favors resonant capture. Therefore, in principle any smaller initial $a_9$ and any $\tau_a$ satisfying the adiabatic criterion should yield similar or larger obliquities than our simulations. We also emphasize that the given migration scheme does not represent the most accurate physical evolutionary pathway of Planet 9's semimajor axis. In reality, Planet 9's outward migration was likely not smooth and its expansion rate likely decreased with time. We will discuss a more realistic migration scheme in Section \ref{sec:stos} - for now, the simple exponential model of semimajor axis growth is selected for ease of use and is sufficient to draw first-order conclusions, given that as long as the evolution remains within the adiabatic limit the precise evolution of the semimajor axis should not significantly impact the dynamics.

Given the uncertainty regarding Planet Nine's orbital parameters, we first analyze results drawn from a wide range of possible orbits

\begin{equation}
    \begin{split}
        m_9 & = 5 - 10 \text{ M}_\Earth \\
        a_9 & = 400 - 800 \text{ AU} \\
        e_9 & = 0.2 - 0.5 \\
        i_9 & = 15^\circ - 25^\circ.
    \end{split}
\end{equation}

\noindent where $m_9, e_9, i_9$ are the initial mass, eccentricity and inclination of Planet Nine in the frame of the ecliptic, respectively. With the context from the wide distribution, we then more closely examine the narrow parameter range suggested by the recent \cite{brown2021orbit} study.

Using draws from the wider range of parameters, we performed a total of 48 dynamical simulations, using the values

\begin{equation}
    \begin{split}
        m_9 & \in \{ 5 \: M_\Earth, 7 \: M_\Earth, 10 \: M_\Earth \} \\
        e_9 & \in \{ 0.3, 0.4, 0.5, 0.6 \} \\
        i_9 & \in \{15^\circ, 20^\circ, 25^\circ, 30^\circ \}.
    \end{split}
\end{equation}

\begin{figure}
    \centering
    \includegraphics[width = 0.45\textwidth]{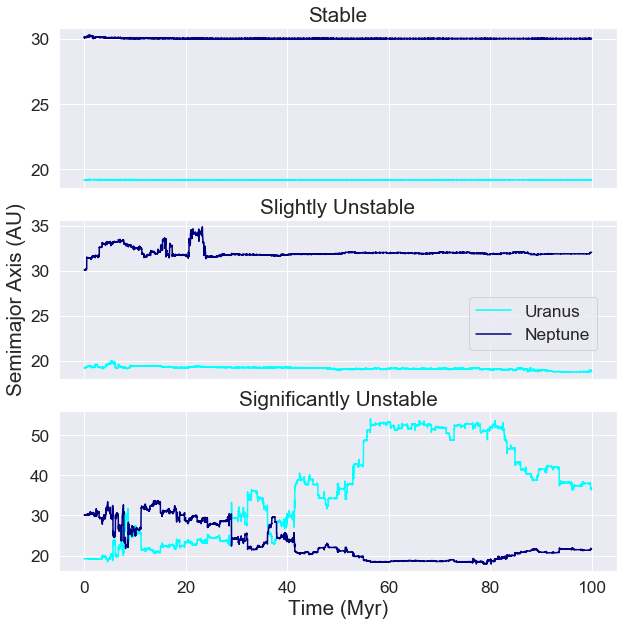}
    \caption{Examples of Uranus and Neptune's semi-major axis evolution for the three stability cases we have defined. The stable case was produced by a Planet Nine with $m_9 = 5$ M$_\Earth$, $e_9 = 0.4, i_9 = 30^\circ$,  the slightly unstable case with $m_9 = 5$ M$_\Earth$, $e_9 = 0.6, i_9 = 20^\circ$, and the unstable case with $m_9 = 7$ M$_\Earth$, $e_9 = 0.4, i_9 = 15^\circ$.}
    \label{fig:stability}
\end{figure}

\noindent Of these 48 simulations, nine resulted in one or more of Uranus, Neptune, or Planet Nine being ejected from the Solar System. These cases are beyond the scope of this paper and are not analyzed further. The remaining 39 simulations exhibit a range of dynamical stability for the orbits of Uranus and Neptune (Jupiter and Saturn remain stable in all cases). We classify them into three categories - \textit{stable}, \textit{slightly unstable}, and \textit{significantly unstable}. Each is defined below (see Figure \ref{fig:stability} for examples of each case).

\begin{itemize}
    \item \textit{Stable} simulations: the semi-major axes of Uranus and Neptune do not diverge from within 10\% of their initial values for the duration of the run. Eighteen of the simulations produced stable outcomes by this measure.
    \item \textit{Slightly unstable} simulations: the semi-major axes of Uranus and/or Neptune diverge by more than 10\% from their initial values, but return to within this threshold by the end of the run. Four simulations are slightly unstable.
    \item \textit{Significantly unstable} simulations: the semi-major axes of Uranus or Neptune differ more than 10\% from their initial values at the end of the run. Seventeen simulations are significantly unstable.
\end{itemize}

\begin{figure*}
    \centering
    \includegraphics[width = 0.95\textwidth]{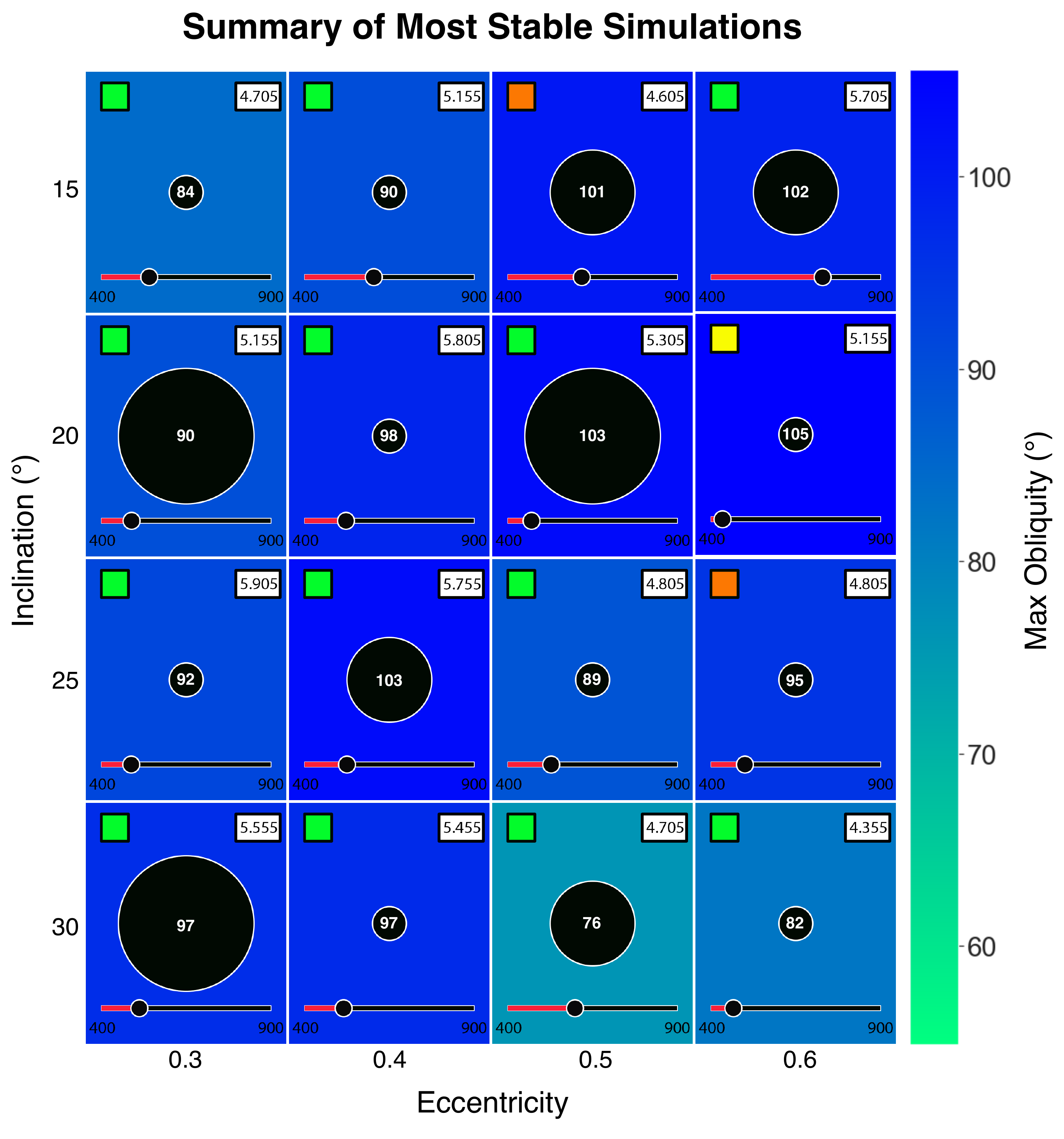}
    \caption{Summary of results of our simulations. Each grid cell contains the most stable result for that eccentricity/inclination pairing. The background color of each grid cell shows the maximum obliquity reached (see colorbar on the right). The size of the central black dot represents the mass of Planet Nine used in that run - small for for 5 M$_\Earth$, medium for 7 M$_\Earth$, and large for 10 M$_\Earth$. The number inside each black dot is the maximum obliquity reached to the nearest degree. The colored square on the upper left hand corner of each grid cell represents the stability of that run - green for stable, yellow for slightly unstable, and orange for significantly unstable. The number in the upper right hand corner of each cell is the value of $\alpha$ used to achieve the maximum obliquity, in arcseconds/year. The slider at the bottom of each cell represents the final semi-major axis of Planet Nine, on a scale from 400 - 900 AU. Three boxes in this plot show yellow or orange stability - this indicates there was no perfectly stable run associated with that cell and the best alternative was selected.}
    \label{fig:gridplot}
\end{figure*}

\noindent Figure \ref{fig:gridplot} summarizes results from our simulations. The format is as follows - each grid cell represents the most stable simulation for an eccentricity/inclination value pair. If two simulations with comparable stability existed for an eccentricity/inclination pairing, the one with the higher maximum obliquity obtained was chosen. Of these simulations, the highest maximum obliquity achieved was $105.6^\circ$, while the lowest maximum obliquity was $75.9^\circ$. The average maximum obliquity reached is $94.7\%$. $13 / 16$, or $81.3\%$, reach a maximum obliquity within $5\%$ of Uranus' present day obliquity of $98^\circ$ or higher, while $6 / 16$ reach a maximum obliquity greater than $98^\circ$. It must be stressed that these high obliquities are reached only with high Uranian precession rates - the associated ramifications will be discussed further in Section \ref{sec:axial}.

\begin{figure}
    \centering
    \includegraphics[width = 0.47\textwidth]{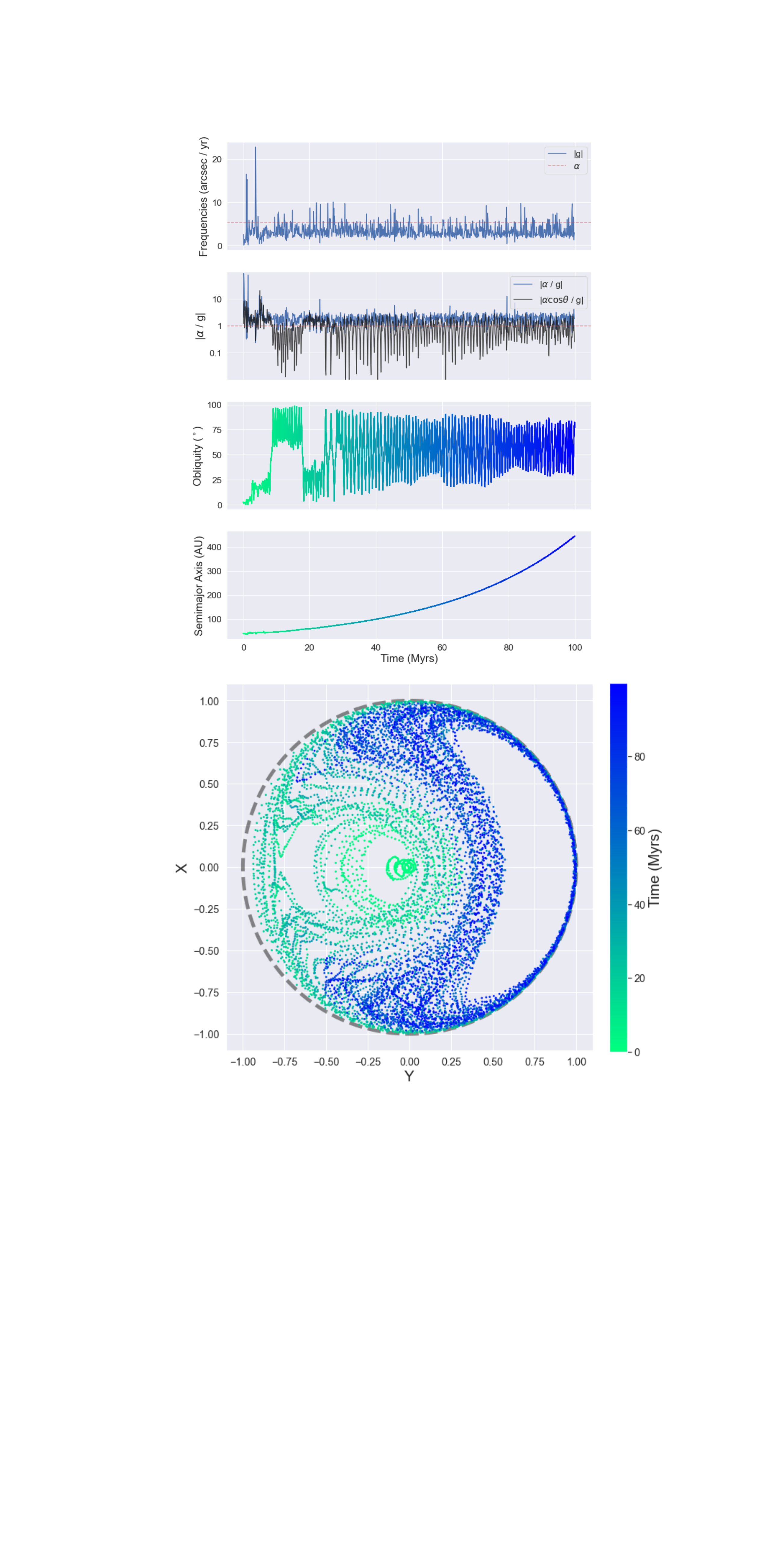}
    \caption{Spin axis evolution for one of our simulations. \textit{Top}: the evolution of $|g|$ (top), the ratio $|\alpha / g|$ (second), obliquity in degrees (third), and Planet Nine's semi-major axis in AU (bottom). The dotted red line in the topmost subfigure marks $|\alpha / g| = 1$. \textit{Bottom}: the evolution of Uranus' spin axis, as calculated with Equation \ref{eq:spin_eom} with $\alpha = 5.305$ arcsec/yr. A maximum obliquity of $103.46^\circ$ is achieved, with Planet Nine's final semi-major axis at 445.4 AU. Note the resonance kicks when the ratio $|\alpha / g|$ crosses unity quickly, and the subsequent resonance capture when the long-term average slowly crosses unity from below.}
    \label{fig:210720ae}
\end{figure}

We now take a closer look at one of the simulations. Figure \ref{fig:210720ae} shows the spin-axis evolution of one of our simulations, as well as the obliquity evolution and Planet Nine's migration trajectory. In this model, we used $m_9 = 10$ M$_\Earth$, $e_9 = 0.5$ and $i_9 = 20^\circ$. Planet Nine's final semi-major axis is 445.39, and we used $\alpha = 5.305$ arcsec/yr to achieve the maximum obliquity value of $103.49^\circ$. Here, we see examples of resonance kicks in both directions (in the 10-30 Myr range) before the spin axis is then captured into resonance and we see the characteristic banana-like librations.

\begin{figure*}
    \centering
    \includegraphics[width = 0.95\textwidth]{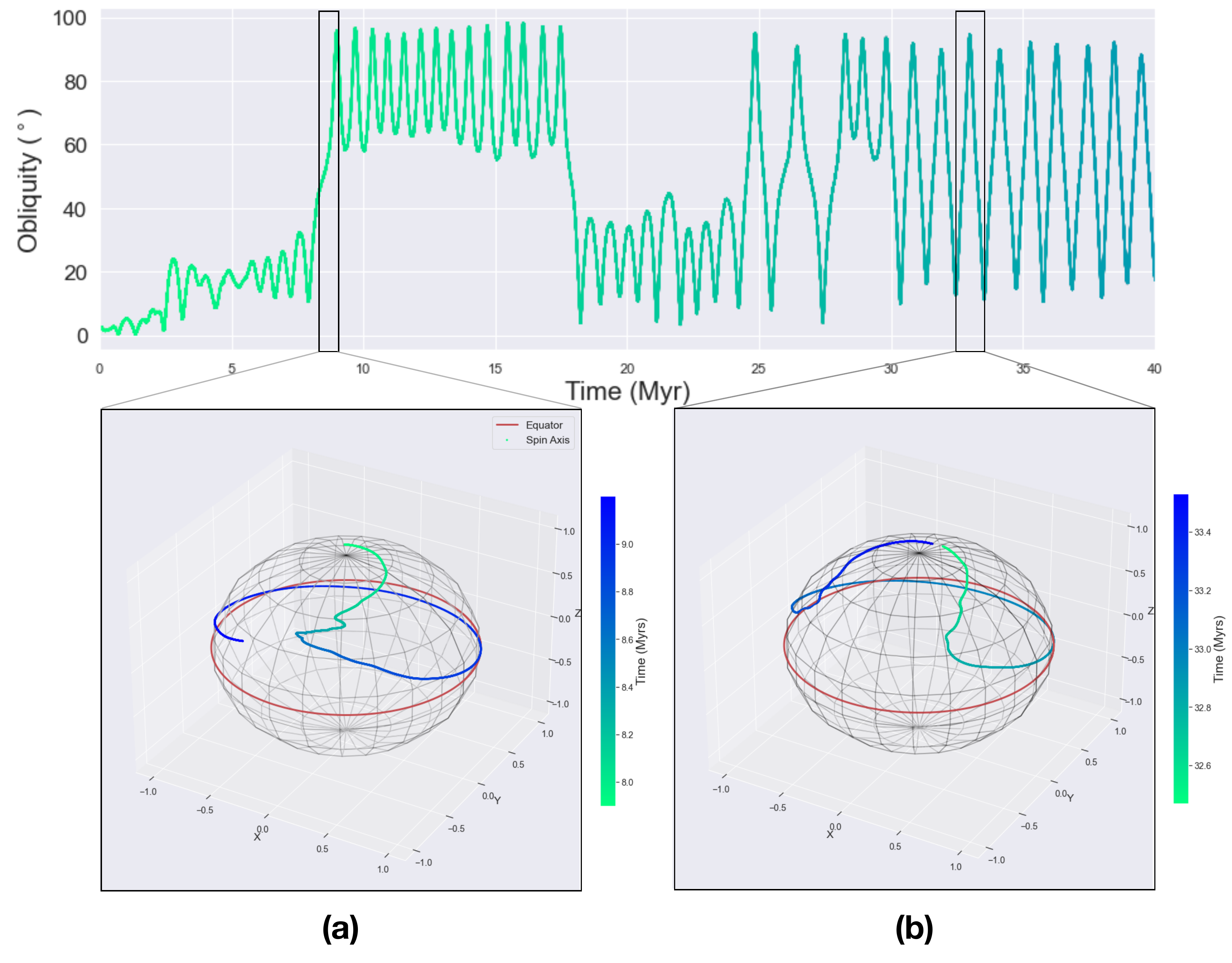}
    \caption{A closer look at two instances where the obliquity of Uranus exceeds $90^\circ$. The top plot shows the first 40 Myr of the obliquity time evolution from Figure \ref{fig:210720ae}. Both bottom figures show 3-D scatterplots of Uranus' spin axis in the designated time slices. The red great circle respresents the projection of Uranus' orbit plane onto the unit sphere, and marks the point at which the obliquity of Uranus exceeds $90^\circ$. In \textbf{(a)}, we see an example of an obliquity kick, as the spin axis continues to circulate after the increase in obliquity. In \textbf{(b)}, the spin axis is shown in resonance, with the characteristic banana-like libration clearly visible.}
    \label{fig:3d_spinplots}
\end{figure*}

A natural question that arises from these simulations concerns our resonance argument: namely, how Uranus achieves obliquities greater than $90^\circ$. From Equation \ref{eq:tau_a}, as the planet's obliquity reaches $90^\circ$ its axial precession rate tends to $0$, breaking the resonance \citep{rogoszinski2020tilting}. However, many of our simulations show maximum obliquities greater than $90^\circ$. \cite{quillen2018tilting} explored a different resonant argument including mean motion terms capable of exciting obliquities above $90^\circ$ - this resonant argument is insensitive to orbital inclination and requires multiple additional planets. We need not appeal to another resonant argument, however - our results can be explained either via obliquity kicks or the amplitude of libration during resonant capture. The simulation shown in Figure \ref{fig:210720ae} is a good illustration of both these cases, as it includes instances of $>90^\circ$ obliquity due to both an obliquity kick and during resonance itself. In Figure \ref{fig:3d_spinplots}, we have zoomed in on a sample of a kick (a), and a resonant libration (b). In (a), note how after the rapid obliquity increase the spin axis fully circulates about the planet, which shows that the spin-axis has not been captured into resonance, and that a rapid kick was responsible for the increase in obliquity. Resonant kicks, which do not satisfy the adiabatic criterion, face no comparable constraint to Equation \ref{eq:tau_a} and thus are capable of exciting obliquities beyond $90^\circ$. In (b), the characteristic banana-like shape of libration is clearly visible - an indication that resonance capture has occurred. Note that obliquities above $90^\circ$ are achieved only at the maximum amplitude of libration, indicating that the libration itself is about a spin-axis position corresponding to an obliquity $\theta < 90^\circ$ hence, there is no contradiction with Equation \ref{eq:tau_a}. The maximum excursions in $\psi$ and $\theta$ from equilibrium over one libration period are related, and can be expressed \citep{hamilton2004}:

\begin{equation}
    \Delta \psi = \sqrt{\tan \theta / \sin i} \Delta \theta
\end{equation}

\subsection{Axial Precession Rate Discrepancy}
\label{sec:axial}
While we have found that Planet Nine is capable of tilting Uranus up to $98^\circ$, there is a caveat to our results - in all cases we require a value of $\alpha$ significantly higher than the present day value. Uranus' current day $\alpha$ is $0.045$ arcsec/yr, two orders of magnitude less than the smallest value used in our simulations. The mechanism of $\alpha$ enhancement has, in the past, been investigated as an attractive avenue for tilting Uranus. \cite{Boue_2010} proposed enhancing the Uranian $\alpha$ with an additional moon, and were able to excite Uranus' obliquity to present day values. This moon, however, would have needed to be very large, with a mass of $0.01$ M$_U$ and a semi-major axis of 50 $R_U$. For reference Uranus' most distant present-day satellite is Oberon with a semi-major axis of $23$ R$_U$, and its most massive satellite is Titania with $m = 4.06 \times 10^{-5}$ M$_U$. This hypothetical moon would raise Uranus' precessional constant to $\alpha = 10.58$ arcsec/yr; however, its size and the fact that it would need to be disposed of at some point raise significant issues with this theory. \cite{rogoszinski2020tilting}, who proposed a circumplanetary disk to enhance $\alpha$, were able to tilt Uranus up to $70$ degrees. In this scenario an impact would still be necessary to reach Uranus' present day obliquity - our simulations have the advantage of requiring no impact at all.

The effect of a circumplanetary disk on a planet's precession rate is well known. For a circumplanetary disk with mass $m_{cp}$, outer radius $r_{cp}$ within the Laplace radius of the planet, and surface density profile $\Sigma_{cp} (r)= \Sigma_{cp, 0} (r/R_p)^{-\gamma}$ \citep{millholland_2019}

\begin{equation}
    \begin{split}
        & q \equiv \frac{1}{2} \left(\frac{2 - \gamma}{4 - \gamma}\right) \left(\frac{m_{cp}}{M_p}\right) \left(\frac{r_{cp}}{R_p}\right)^2 \\
        & l \equiv \left( \frac{2 - \gamma}{ 5 / 2 - \gamma} \right) \left( \frac{m_{cp}}{M_p R_p^2 \omega} \right) \left(G M_p r_{cp}\right)^{1/2},
    \end{split}
\end{equation}

These values are inserted into Equation \ref{eq:alpha} to calculate the corresponding Uranian $\alpha$. It is thus useful to assess the feasibility of our results in the context of a circumplanetary disk. \cite{Szul_gyi_2018} performed radiative hydrodynamic simulations to estimate a reasonable initial mass for the Uranian circumplanetary disk of $\sim 7.4 \times 10^{-4}$ M$_U$. Using this value and taking a fiducial value for the circumplanetary disk radius of $r_{cp} = 54$ R$_U$ based on the Laplace radius \citep{rogoszinski2020tilting}, we arrive at $\alpha = 1.06$ arcsec/yr. However, \cite{Szul_gyi_2018} shows that the circumplanetary disk is not a closed system but rather is continuously fed mass by the circumstellar disk at a rate of $2 \times 10^{-3}$ M$_U$/yr. Given this, a larger disk of $m_{cp} = 2.9 \times 10^{-3}$ M$_U$ and $r_{cp} = 75$ R$_U$ can reasonably be assumed, which is sufficient to give $\alpha = 6$ arcsec/yr, which encompasses the range of all our simulations. \citep{rogoszinski2020tilting} argue that the traditional Laplace radius of a planet is potentially enhanced by a factor of four in the presence of a circumplanetary disk, so this larger required radius is reasonable and fits within the potential Laplace radius of the planet. These values consider a constant surface density disk - for a disk with surface density gradient $\gamma = 3/4$ \citep{canup_2002, millholland_2019}, a disk mass of $m_{cp} = 4.1 \times 10^{-3}$ M$_U$ is required.

Our brief circumplanetary disk analysis oversimplifies a deep and richly complex field. \cite{mamajek_2012} estimate that the lifetime of a circumplantary disk at Uranus' present-day orbit is of order 10 Myr - a timeframe in modest tension with the durations of our simulations. The possibility of dips in the circumplanetary disk's surface density near the Laplace radius and detachment from the ecliptic plane at high obliquities have also been noted by \cite{rogoszinski2020tilting} and \cite{tremaine_2014}, respectively - the latter scenario is of particular concern, as detachment from the ecliptic would shrink the rate of precession and potentially break the resonance. Neither of these effects are considered in this argument. Finally, we also do not consider the effect of Uranus accreting from the circumplanetary disk over time - as Uranus accretes matter and gains angular momentum, from Equation (\ref{eq:alpha}) all else being equal the axial precession rate $\alpha$ will decrease \citep{rogoszinski2020tilting}. This works against the increase of $|\alpha / g|$ - resonance capture will occur more slowly. If we are to account for accretion from the circumplanetary disk, we would expect longer timescales to reach the obliquities in this paper, dependant on the rate of accretion from the disk. We emphasize that we are agnostic with regards to the means of this necessary precessional frequency enhancement and offer this discussion of the effect of circumplanetary disks to provide an order-of-magnitude feasibility analysis for one of several viable options.

\subsection{The Latest Planet Nine Parameters}

We now present results from a set of simulations more closely analyzing the parameter space predicted by \cite{brown2021orbit} (see Equation \ref{eq:p9params}). Again, we initialize each REBOUND simulations with the four outer planets, and Planet Nine at 40 AU. We set $\tau_a = 4.09 \times 10^7$ years and integrate for $10^8$ years. This in principle gives

\begin{equation}
    a_f = a_i e^{t_{max} / \tau_a} = 460 \text{ AU},
\end{equation}

\noindent though again effects of the other bodies in the system make the true final semi-major axis ultimately unpredictable. We ran 75 simulations with the values

\begin{equation}
    \begin{split}
        m_9 & \in \{ 4.9 \: M_\Earth, 6.2 \: M_\Earth, 8.4 \: M_\Earth \} \\
        e_9 & \in \{ 0.20, 0.25, 0.30, 0.35, 0.40\} \\
        i_9 & \in \{11^\circ, 13^\circ, 16^\circ, 18^\circ, 21^\circ \}.
    \end{split}
\end{equation}

\begin{figure*}
    \centering
    \includegraphics[width = 0.95\textwidth]{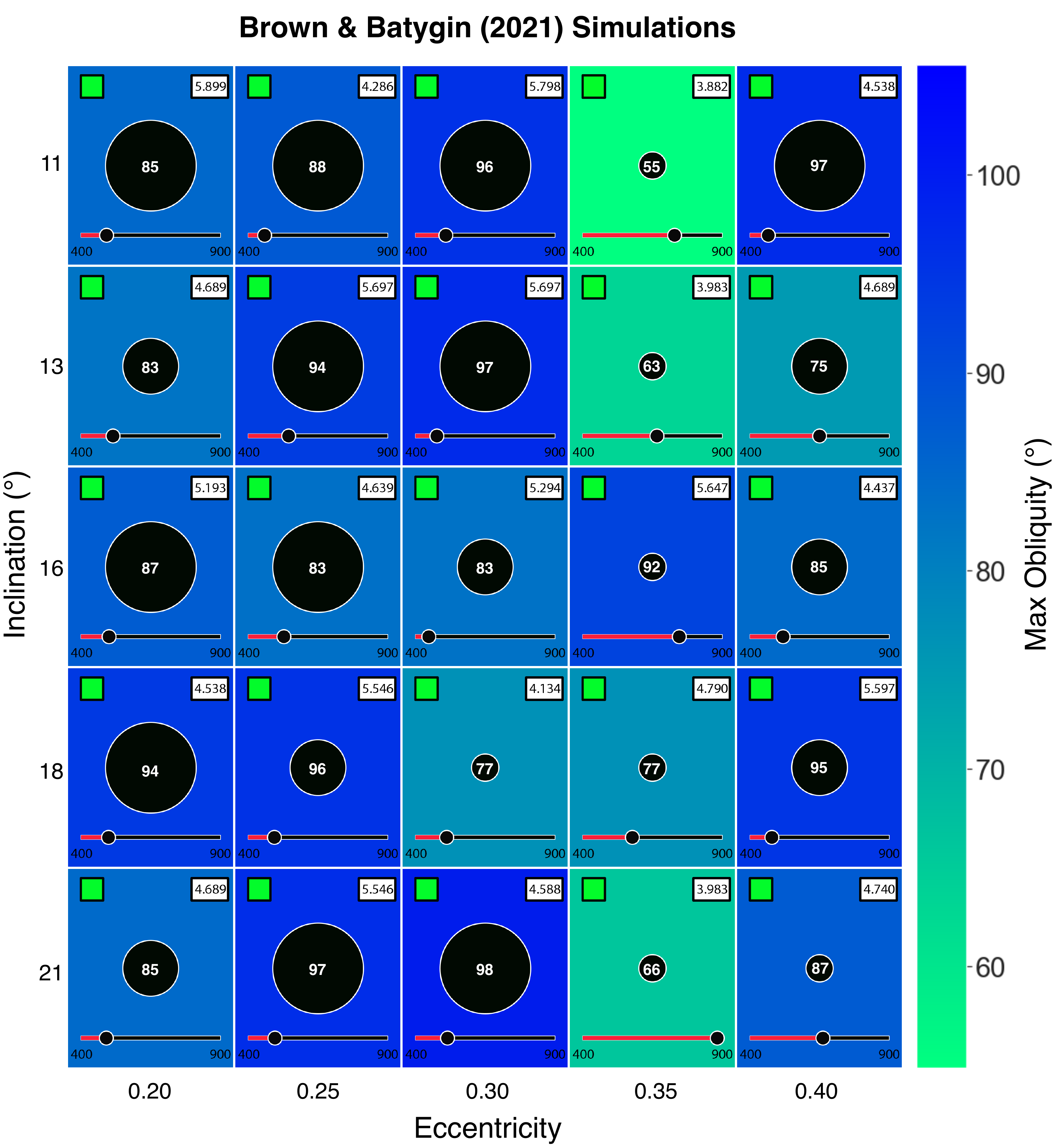}
    \caption{Summary of results of our most stable simulations with the updated parameters from \cite{brown2021orbit}. Quantities are as described in Figure \ref{fig:gridplot}, with the exception of mass - a small central circle for $m_9 = 4.9$ M$_\Earth$, a medium one for $m_9 = 6.2$ M$_\Earth$, and a large one for $m_9 = 8.4$ M$_\Earth$.}
    \label{fig:gridplot_small}
\end{figure*}

\noindent Two of these simulations resulted in a body being ejected and were not considered in our analysis. Of the remaining 73, 53 were stable, 8 were slightly unstable, and 12 were significantly unstable. Figure \ref{fig:gridplot_small} summarizes these results in the same format as Figure \ref{fig:gridplot}. On the whole, maximum obliquities tend to be lower than the more general case; nevertheless, $10/25$ (or $40 \%$) of the simulations reach a maximum obliquity within $10\%$ of Uranus' present day value. This round of simulations favors the most massive $m_9 = 8.4$ M$_\Earth$ case.

\subsection{Stochastic Scattering}
\label{sec:stos}
As a brief caveat, note that simple exponential migration as we have modelled does not necessarily represent the most physically accurate pathway for Planet Nine's current orbit. Rather, as a typical TNO is scattered outwards its perihelion distance is coupled to Neptune and remains constant, while the semimajor axis grows stochastically. For such a stochastic system, the dynamics of semimajor axis evolution may be described by a conventional diffusion equation, where the relevant physics are encapsulated by the diffusion coefficient \citep{batygin_mardling_2021}

\begin{equation}
    \mathcal{D}_a = \frac{8}{5 \pi} \frac{m_P \sqrt{G M_\odot a_P}}{M_\odot} \exp \left[ -\frac{1}{2} \left(\frac{q}{a_P}\right)^2\right].
\end{equation}

Here $m_P, a_P$ are the mass and semimajor axis of the coupled planet (for a typical TNO this is Neptune - in the case of Planet 9's migration Saturn is more appropriate) and $q$ is the perihelion of Planet Nine.

\begin{figure}
    \centering
    \includegraphics[width=0.5\textwidth]{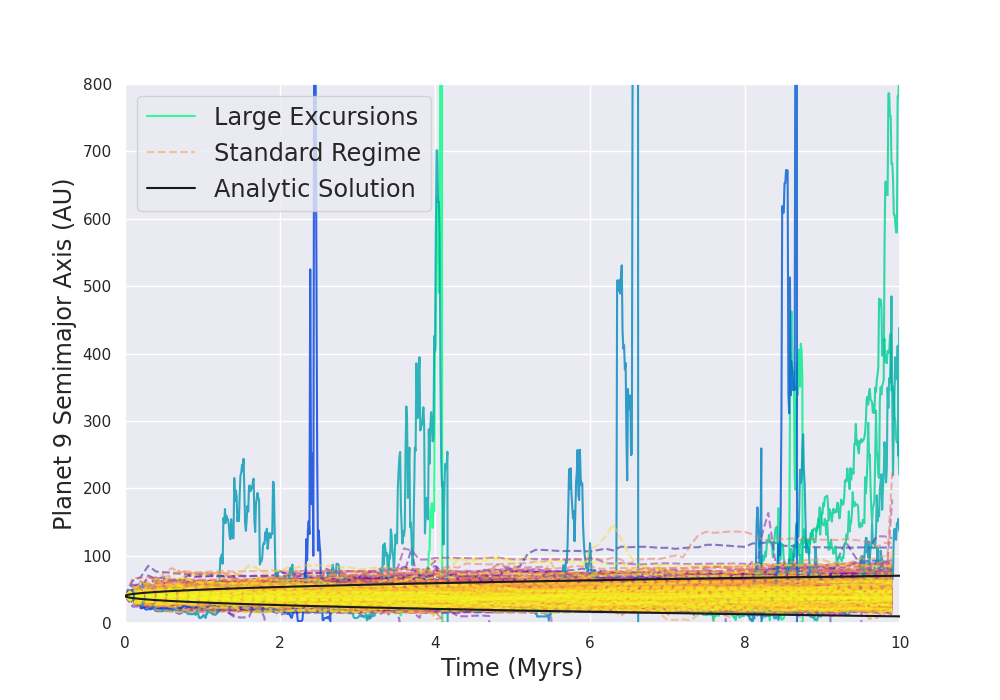}
    \caption{400 self-consistent REBOUND simulations of Planet 9's dynamical migration. Each line corresponds to one simulation - in each one, Planet 9 is initialized with a random phase. The yellow-purple lines represent pathways which do not deviate far from the expected analytic solution (in black), while the blue-green lines represent pathways which show large excursions from the mean.}
    \label{fig:diffusion}
\end{figure}

As previously mentioned \cite{Bromley_2016, li2016interaction, Eriksson_2018}, the issue of Planet 9's migration is a difficult one. We have performed a suite of numerical simulations as our own brief analysis of the issue. Figure \ref{fig:diffusion} shows the results of 400 REBOUND simulations integrated over 10 Myr. Each simulation is initialized as follows - the four giant planets in their present-day configuration, and Planet 9 initialized at 40 AU with a perihelion of 25 AU ($e_9 = 0.375$) and a random phase. No additional forces are imposed - the resulting migration of Planet 9 is fully self-consistent and arises purely from perturbations from the other planets. The shaded region of the plot represents the analytic solution of \cite{batygin_mardling_2021}: in time $t$ the expected scattering is $\pm \sqrt{\mathcal{D}_a t}$. For the vast majority of simulations, the numeric results are in good agreement with the analytic prediction. The large excursions from the norm are the cases relevant to our study: these would be the pathways resulting in a Planet 9 which would match present-day parameters. It is clear that an instability which would both match Planet 9's present-day configuration and occur on a timescale slow enough to enter secular-spin orbit resonance is very difficult to reproduce. For this reason, a fully self-consistent picture of Planet 9's migration and its effect on the Uranian spin axis is beyond the scope of this paper. We note that these large excursions need not be in conflict with Uranus' large present-day obliquity - we can imagine a scenario where Planet 9 diffuses in the "standard" smooth migration regime sufficiently long to excite Uranus' obliquity, and then subsequently enters the regime of large excursions to place it at its present-day location.

\section{\textbf{Discussion}}
\label{sec:disc}
We have computed an array of simulations spanning Planet Nine's parameter space that are capable of exciting Uranus' obliquity to very high values. From the set of simulations encompassing the most general Planet Nine parameter space from \cite{Batygin_2019}, $81.2 \%$ are able to reach or exceed within $10\%$ of the present-day value of $98^\circ$. We also present an array of simulations more closely examining the parameter space given by the most recent Markov Chain Monte Carlo analysis \citep{brown2021orbit}, and find that $40\%$ of these simulations are able to reach within $10\%$ of Uranus' present day obliquity. Given our results, we conclude that it is possible to tilt Uranus over via a spin-orbit resonance mechanism driven by Planet Nine's outward migration. The feasibility of this scenario is not as straightforward. While we are, in several cases, able to tilt Uranus to $98^\circ$, we require an axial precession rate $\alpha$ two orders of magnitude greater than Uranus' present day $\alpha = 0.045$ arcsec/yr, and a few times larger than the disk sizes favored by the hydrodynamic simulations of \cite{Szul_gyi_2018}. We also imposed a smooth exponential migration scheme, with is not the most realistic migration pathway for Planet 9. The feasibility of our results thus depends on the probability of a more robust $\alpha$ enhancement, perhaps through larger circumplanetary disks \citep{rogoszinski2020tilting} or a primordial moon \citep{Boue_2010} - more work in assessing the potential for Uranus' primordial precession rate, as well as Planet 9's precise migration history, will be vital in assessing our hypothesis.

This work only considers simulations which resulted in reproducing the Solar System as it stands today, and heavily favored systems that exhibited a high degree of stability. In fact, we did produce several unstable simulations which were able to drive extremely high obliquities (up to a maximum of $134.16^\circ$), but were rejected due to a high degree of instability or even the ejection of a planet. Of course reproducing the Solar System's current state is paramount, but instability does not necessarily preclude this possibility. In fact, the Nice model \citep{tsiganis2005, gomes2005, morbi2005} both reproduces the present day Solar System and predicts migration in each of the giant planets, with some models admitting the possibility of a planet's ejection \citep{Batygin_2011}. \cite{vokrouhlicky2015tilting} have shown that the outward migration predicted by the Nice model is consistent with Jupiter and Saturn's present day axial tilts. \cite{rogoszinski2021tilting} have taken significant strides in investigating Uranus' obliquity evolution within the context of planetary migration (in their case, Neptune) - further work in placing Planet Nine's migration in this greater framework has the potential to yield interesting results, both in general and applied to Uranus' axial tilt.

The implication that planetary migration can induce high obliquity in other planets of the system is an intriguing one. Most notably, the fact that we are able to tilt Uranus to such an extent raises the question of Neptune's $30^\circ$ obliquity and if Planet Nine's migration could be responsible as well. More broadly, obliquity plays an important role in assessing the viability of an exoplanet's habitability, and direct measurement has proven difficult \citep{Shan_2018}. With the recent detection of a circumplanetary disk by \cite{Benisty_2021}, it seems feasible that more robust estimates of exoplanet $\alpha$ values may be forthcoming.

\acknowledgments
We thank Konstantin Batygin, Garrett Levine, Harrison Souchereau, Tanvi Gupta and our colleagues at the Yale Exoplanet Journal Club for useful feedback and discussions. We also thank the anonymous reviewers for helpful comments that greatly improved the manuscript. T.L would like to thank the Yale Center for Research Computing (YCRC) for valuable support and resources over the course of the project.

\newpage

\bibliography{sample63}{}
\bibliographystyle{aasjournal}

\end{document}